\documentclass[aps,twocolumn,prl]{revtex4}
\usepackage{amsfonts}
\usepackage{epsf}
\usepackage{graphicx}
\usepackage{amsmath}
\usepackage{amssymb}
\usepackage{amstext}

\begin{document}

\title{Nano-Engineering Defect Structures on Graphene}
\author{Mark T. Lusk and Lincoln D. Carr}
\affiliation{Department of Physics, Colorado School of Mines, Golden, CO 80401, USA}

\begin{abstract}
We present a new way of nano-engineering graphene using defect domains.
These regions have ring structures that depart from the usual honeycomb
lattice, though each carbon atom still has three nearest neighbors. A set
of stable domain structures is identified using density functional theory
(DFT), including blisters, ridges, ribbons, and metacrystals.  All such
structures are made solely out of carbon; the smallest encompasses just 16
atoms.  Blisters, ridges and metacrystals rise up out of the sheet, while
ribbons remain flat.  In the vicinity of vacancies, the reaction barriers
to formation are sufficiently low that such defects could be synthesized
through the thermally activated restructuring of coalesced adatoms.
\end{abstract}

\maketitle

Carbon is a fundamental material for nano-engineering.  Three dimensional
graphite, one-dimensional nanotubes, zero-dimensional buckyballs, and now
two-dimensional graphene are all being investigated intensely to this end.
The latter, recently created experimentally for
the first time~\cite{novoselov2005,zhangYB2005},
is the first
stable two-dimensional solid state lattice material.
Fundamental aspects of graphene include the room-temperature quantum Hall
effect and an effective description by a Dirac equation~\cite{geim2007}.
Graphene has significant applications in electronics~\cite{schedin2006,silvestrov2007,green2007},
and is even predicted
to replace silicon in future solid state devices~\cite{geim2007}.
Many of these applications
require cutting sheets of graphene precisely in order to localize charge,
among other effects.  We propose an alternative method of patterning
graphene without cuts of any kind, which can be adapted to many
applications. The key to this method is controlled placement of
groups of defects,
called \emph{defect domains}.

\begin{figure}[hptb]\begin{center}
\includegraphics[width=0.45\textwidth]{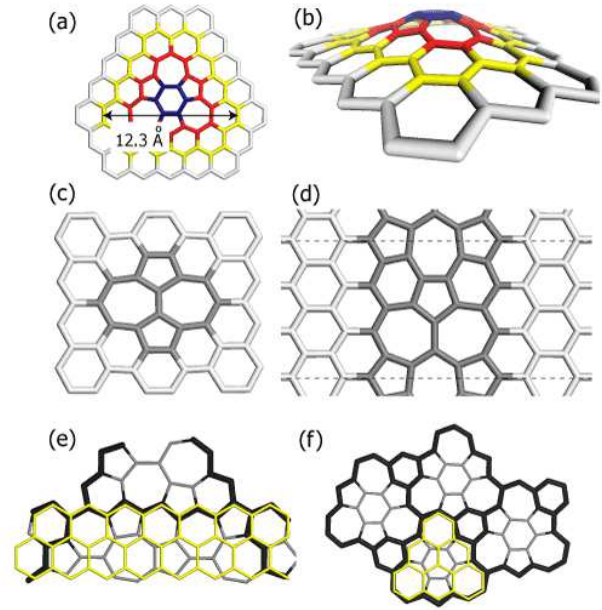}
\caption{%
Defect Domains in Graphene. (a)-(b) Top and side view of a single defect
blister of 24 carbon atoms with a height of 1.9 \AA . (c) A
Stone-Wales defect used to synthesize defect domains. (d) A defect domain
ribbon embedded in graphene. (e)-(f) Haeckelite, a carbon allotrope used to
synthesize defect domains. In yellow is shown a graphene overlay for
visualization.}\label{tile_graphic}%
\end{center}\end{figure}

In this Letter, we demonstrate that a
complex, stable landscape of defect domains can be monolithically nano-engineered
from graphene to make arbitrary structures.
An example of a defect domain which stands up out of the graphene
plane is shown in Fig.~\ref{tile_graphic}(a)-(b).  Such \emph{defect blisters} in graphene
can be generated through the
coalescence of adatoms in the vicinity of vacancies (Fig.~\ref{tile_graphic}(c)). \emph{Defect ridges} can
be constructed as an alignment of blisters and their two-dimensional
patterning results in \emph{defect metacrystals}.  Planar, \emph{defect
ribbons} may also be synthesized as shown in Fig.~\ref{tile_graphic}(d).
These structures, in particular, may be viewed as being spliced-in pieces of
a carbon allotrope known as haeckelite~\cite{terrones2000}.  Figure~\ref{tile_graphic}(e)-(f)
show how haeckelite symmetry considerations may be
used to predict energetically favorable defect domain geometries.
We use density functional theory (DFT) to study stability, scalability, and
other key physical properties of these materials.  DFT has been applied
successfully to magnetism in graphene
nano-islands~\cite{fernandezJ2007}, the effect of substrates~\cite{varchonF2007},
and transport in doped graphene nanoribbons~\cite{martins2007};
DFT is also used for carbon nanotubes~\cite{chandrasekera2007, borstnika2005} and
other carbon structures~\cite{hamm2007}.

Defects have been observed in graphene~\cite{geim2007,hashimoto_2004}
and are expected to play a key role in the functional
properties of working materials~\cite{lu_2004}.  Among the simplest and more
easily fabricated of these is the Stone-Wales defect, wherein the rotation
of a single pair of carbon atoms creates adjacent pairs of pentagonal and
heptagonal rings~\cite{stone_1986}, as shown in Fig.~\ref{tile_graphic}(c).
Such defects can be introduced intentionally using electron radiation and
imaged using Transmission Electron Microscopy (TEM)~\cite{hashimoto_2004}.
Electron radiation also causes the formation of adatom/vacancy pairs which
subsequently separate and move across the graphene lattice~\cite
{hashimoto_2004}. DFT analysis~\cite{krasheninnikov2004} estimates the
adatom migration barrier to be 0.45 eV with a jump frequency of $(3.7 \pm
0.7) \times 10^{12}\: \mathrm{s}^{-1}$, while the barrier to vacancy migration is estimated to be
1.7 eV with a jump frequency on the same order as for adatoms~\cite{elbarbary2003}.
Electron radiation can therefore be used to generate
adatoms and vacancies which have been observed to rearrange themselves to
form more complex defect structures~\cite{hashimoto_2004}.

Domains over which similar defect structures are periodically replicated are known as
haeckelite~\cite{terrones2000}, as shown in Fig.~\ref{tile_graphic}(f).
Tight binding molecular dynamics studies suggest that small regions of these
carbon allotropes may be induced within graphene through the coalescence of
four single vacancies~\cite{lee_2006}. In principle, entire sheets and
tubes of haeckelite can be synthesized which, even without experimental
realization, have already generated potential applications~\cite{mpourmpakis_2006}.
We focus on the simpler notion of localized defect structures
which are minimally extended beyond Stone-Wales defects.
We initially treat defects composed of ribbons and patches of
haeckelite and then consider a new fundamental building block which we
term an \emph{Inverse Stone-Wales defect}.

All calculations were performed with the real-space, numerical atomic
orbital, DFT code, DMOL~\cite{JChemPhys.113.7756}. A norm conserving, spin
unrestricted, semi-core pseudopotential approach was employed with electron
exchange and correlation accounted for using the Perdew-Wang generalized gradient
approximation (GGA)~\cite{PhysRevB.45.13244}. Periodic boundary conditions
 were employed and vacuum slabs were used to isolate the replicated
graphene layers.  As a check on the
method, the ground state energy of C$_{60}$, i.e., a buckyball, was
estimated to be 384 meV/atom above that of graphene, consistent with a
literature value of 380 meV/atom~\cite{terrones2004}. Likewise, a single
Stone-Wales defect was estimated to have a formation energy of 5.08 eV when
embedded within a 144-atom graphene supercell; this compares well with
an estimate of 4.8 eV from the literature~\cite{li:184109}.

\begin{figure}[ptb]\begin{center}
\includegraphics[width=0.45\textwidth]{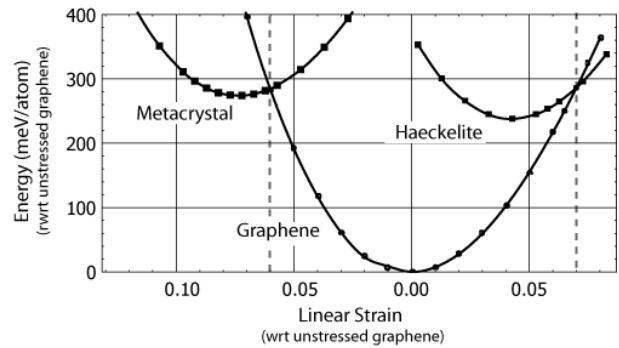}
\caption{Mechanical response (pure dilation) of graphene, haeckelite, and a
metacrystal of haeckelite-based defect domain blisters.  The points represent
actual DFT data; the curves are a guide to the eye.}
\label{stability_exchange_2}%
\end{center}\end{figure}

We begin with defect structures constructed from patches of
haeckelite~\cite{terrones2000}.  The H$_{5,6,7}$ variant with hexagonal symmetry
was chosen over rectangular and oblique allotropes.  This is the most stable
of the three variants.  We find a ground state energy estimate for H$_{5,6,7}$ of 229
meV/atom above graphene.  This is close to the literature value of 246
meV/atom and notably lower than that of C$_{60}$~\cite{terrones2004}.
H$_{5,6,7}$ is $3\%$ less dense than graphene: it has 0.369 atoms/\AA$^{2}$,
compared to 0.380 atoms/\AA$^{2}$ for graphene. Haeckelite can therefore be formed by the removal of carbon atoms or in settings in which the graphene sheet can expand. For instance, a 20-atom periodic cell of this carbon allotrope can be formed through the addition of two carbon atoms to a Stone-Wales defect followed by lateral dilation. The energy barrier is estimated to be 1.8 eV per cell. Such restructuring should be aided by straining the graphene.  DFT calculations support this and indicate that the
stability exchange occurs at a linear strain of $\sim$0.07,
as shown in Fig.~\ref{stability_exchange_2}.
The analysis also determined uni-axial elastic
constants for graphene and haeckelite of 1.13 GPa and 1.20 GPa,
respectively.  The first number is consistent with experimental measurement
of 1.06 GPa~\cite{PhysRevB.7.4527}, while the second indicates that haeckelite
is 6\% stiffer than graphene.

Two types of approximately coherent interfaces between graphene and H$_{5,6,7}$
can be identified by inspection and are shown in
Fig.~\ref{tile_graphic}(e)-(f).  A linear haeckelite strain of $4.7\%$ gives
averaged coherence along the symmetry boundaries shown in Fig.~\ref{tile_graphic}(e),
while triangular patches of haeckelite
are approximately coherent without applying any dilation, as illustrated in Fig.~\ref{tile_graphic}(f).
With these lines as a
guide, haeckelite can be spliced into a graphene sheet and vice-versa.
A strip of haeckelite spliced into graphene along adjacent, linear coherency
lines has a planar ground state, as shown in Fig.~\ref{tile_graphic}(d),
where the dashed lines indicate the periodicity of our simulation.  On the
other hand, splicing a triangular tile from Fig.~\ref{tile_graphic}(f) into
graphene causes a ground state blister to form, as shown in Fig.~\ref{tile_graphic}(a)-(b). The converse, a triangular graphene patch within haeckelite, results in a planar structure.

\begin{figure}[ptb]\begin{center}
\includegraphics[width=0.45\textwidth]{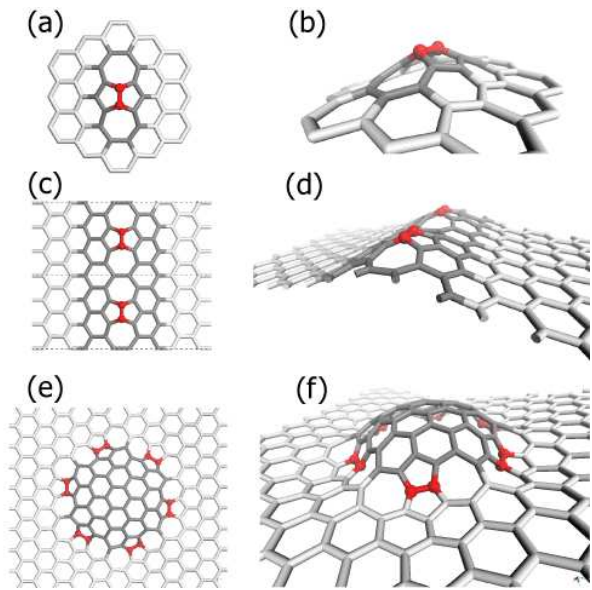}
\caption{
(a)-(b) A single Inverse Stone-Wales defect.  (c)-(f) Multiple
defects can be aligned to form ridged contours
which can be straight, curved, or even closed.
}
\label{isv_collage}
\end{center}\end{figure}

The concept of a defect blister creates a natural way to extend graphene out of the
plane.  A formation energy of 6.07 eV was determined for the
blister illustrated in Fig.~\ref{tile_graphic}(a)-(b), using a
202-atom supercell. This is approximately 1 eV higher than the Stone-Wales
defect. A Hessian analysis~\cite{wilson1955} on a smaller supercell
indicates that such haeckelite-based defect blisters are linearly stable.  This is
consistent with the stability observed in a room temperature quantum
molecular dynamics simulation, which we also performed.

We now turn to an even simpler defect domain structure not based on Haeckelite,
and explicitly describe its reaction pathway.
While haeckelite-based defect blisters amount to the substitution of a pair of atoms for
a hexagonal ring (Fig.~\ref{tile_graphic}(f)), a narrower structure can be
formed by inserting two atoms on opposing faces of an existing hexagon as
shown in Fig.~\ref{isv_collage}(a)-(b).  This results in pairs of pentagons
and heptagons that we refer to as an Inverse Stone-Wales defect
(compare to Fig.~\ref{tile_graphic}(c)). A 16-atom blister results with a
footprint of $12.2 \times 7.4$ \AA  and a height of 2.1 \AA.
A formation energy of 6.20 eV was determined using a 200-atom supercell,
nearly the same energy as that for the larger haeckelite
blister. A series of Inverse Stone-Wales blisters can be aligned to form a
corrugated ridge of nearly arbitrary contour (Fig.~\ref{isv_collage}(c)-(d)),
a monolithic analog of carbon nanotubes formed from C$_{60}$
molecules~\cite{rodriguez2004}.  Such structures may hold utility in guiding
charge transport on graphene.  Further, the creation of closed contours
results in extended blisters of pure graphene enclosed by a ring of Inverse Stone-Wales defects.
(Fig.~\ref{isv_collage}(e)-(f)).

\begin{figure}[ptb]\begin{center}
\includegraphics[width=0.45\textwidth]{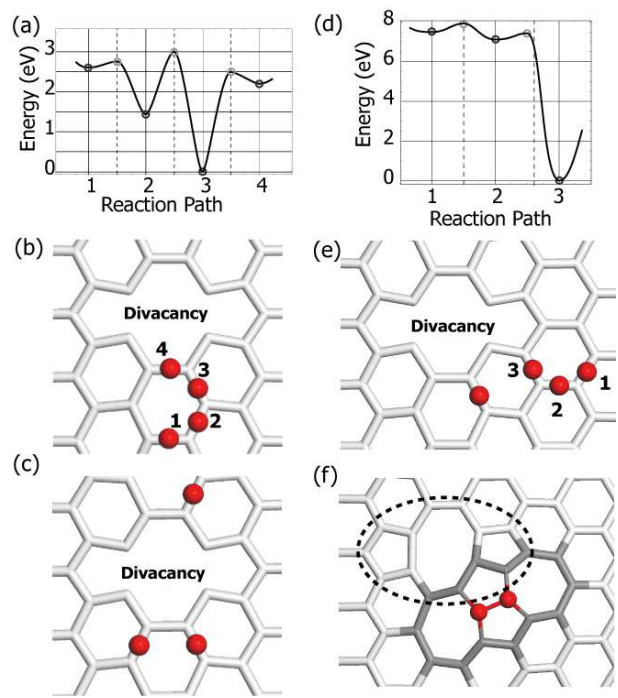}
\caption{
(a)-(b) Reaction path for a single adatom moving to the periphery of a
di-vacancy. (c) Experimentally observed adatom positions~\cite{hashimoto_2004}.
(d)-(e) Creation of an Inverse Stone-Wales defect on the periphery of a di-vacancy
by one adatom hopping within the vicinity of another. (f) Final structure
with blister shown in dark gray and di-vacancy identified within a dashed ellipse.
Energies in both (a) and (f) are relative to the lowest point plotted.}
\label{reaction_paths}%
\end{center}\end{figure}
The synthesis of defect structures may be facilitated by electron radiation
and the resulting collection of adatoms and vacancies~\cite{hashimoto_2004}.
The formation energy for a single vacancy was calculated to be 7.63 eV,
consistent with experimental estimates~\cite{thrower1978},
and therefore has the potential to provide a large driving force for restructuring
events within its horizon.  The formation energy of a vacancy/adatom pair, the planar analog of a Frenkel defect, is 14.13 eV.  Di-vacancy formation energy was calculated to be 8.08 eV. In addition, re-structuring within such a low dimensional system is
more easily carried
out than in bulk since vacancies facilitate significant
low energy distortions of the local
lattice. To explore this, a hybrid linear synchronous transit/quadratic
synchronous transit (LST/QST) transition state search algorithm~\cite{govind2003}
was used to construct a reaction pathway for the approach
of a single adatom to a di-vacancy (Fig.~\ref{reaction_paths}).  Adatoms hop
between adjacent bridge sites, and a barrier of 0.52 eV was identified for
jumps in pure graphene, consistent with an estimate of 0.45 eV obtained
elsewhere~\cite{krasheninnikov2004}.  However, the di-vacancy attracts adatoms to
positions that are one hop removed from its periphery (position 3 in Fig.~\ref{reaction_paths}(b)),
and the associated barriers can be overcome by
thermal fluctuations at modest temperatures. These are precisely the
bridge sites at which adatoms are observed experimentally
(Fig.~\ref{reaction_paths}(c))~\cite{hashimoto_2004}.
The reaction energies in Fig.~\ref{reaction_paths}(a)
are $E^r_{12} = -1.17$ eV, $E^r_{23} = -1.43$ eV, and
$E^r_{34} = 2.20$ eV, while the reaction barriers are $E^b_{12} = 0.14$ eV,
$E^b_{23} = 1.55$ eV, and $E^b_{34} = 2.49$ eV.

\begin{figure}[ptb]\begin{center}
\includegraphics[width=0.45\textwidth]{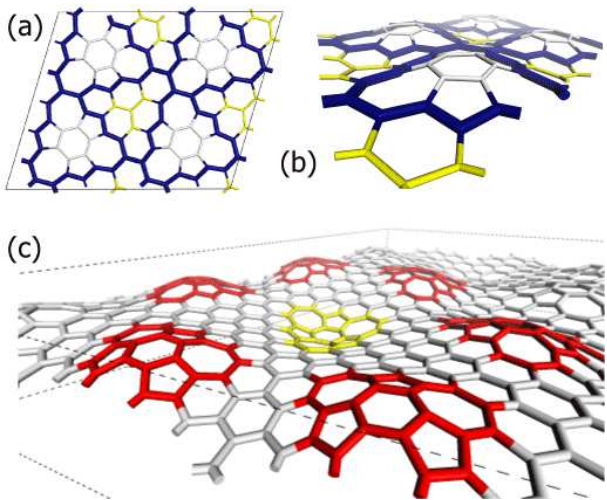}
\caption{
(a)-(b) Top and side view of a metacrystal formed
from haeckelite-based defect-domain blisters.
(c) A metacrystal defect: a less closely packed metacrystal contains an ``atom'' of negative polarity,
a blister pushed down rather than up.
}
\label{mini_cp_graphic}
\end{center}\end{figure}
Two adatoms are required to synthesize an Inverse Stone-Wales blister.
So a second transition state analysis was carried out wherein one adatom
jumped towards another located at a low energy site on the periphery of the
di-vacancy, as shown in Fig.~\ref{reaction_paths}(e)-(f).
As indicated in the figure,
the reaction barriers can be overcome by thermal fluctuations at modest
temperatures and result in a blister. The reaction energies are
$E^r_{12} = -0.40$ eV and $E^r_{23} = -7.10$ eV, while the reaction barriers
are $E^b_{12} = 0.39$ eV and $E^b_{23} = 0.34$ eV. Subsequent vacancy
migration or elimination would leave the stand-alone blister shown in
Fig.~\ref{isv_collage}(a)-(b).

Localized defect domains are of interest in their own right, but
patterned arrangements of them may endow graphene sheets with novel
metacrystalline properties. Within such a paradigm, blisters represent
meta-atoms with either up or down polarity. A mono-sized, mono-polarity,
ground state, metacrystal is shown in Fig.~\ref{mini_cp_graphic}(a)-(b).
The associated mechanical response given in Fig.~\ref{stability_exchange_2}
indicates that it has a density
greater than that of graphene. The ground state energy is only 274
meV/atom above that of graphene and is surprisingly close to the ground
state energy of pure haeckelite, which is 229 meV/atom above graphene.
This suggests that the metacrystals are more energetically stable than C$_{60}$;
the latter has an analogous energy of 384 meV/atom. The elastic stiffness of the
metacrystal is 1.08 GPa in uni-axial tension, somewhat less that either
graphene or haeckelite.  One possibility for the experimental realization
of defect metacrystals in graphene is interference patterns of electrons in the near
field of a diffraction grating~\cite{croninAD2006}. Such a system has been
demonstrated at the 100 nm scale already and has the potential to be reduced
to the 10 nm scale. Metacrystals can also be made with larger
blisters or even blisters of alternating size.  Thus one can consider a
``periodic table'' of meta-atoms, creating a variety of 2D
metacrystals of arbitrary lattice and crystal structure.

In conclusion, we have described a new method to achieve monolithic
nano-engineering of graphene via defect domains.
These structures can take the form of blisters, ribbons, and ridges.
We showed that such objects are linearly stable and can be arranged
in arbitrary patterns, leading to a metacrystal.
The smallest defect domain,
referred to as an Inverse Stone-Wales defect, consists of only 16 atoms, and
the associated energy is approximately 1 eV higher than a simple
Stone-Wales defect.  Di-vacancies were shown to attract adatoms to their
periphery and a thermally activated path was identified for blister
synthesis.
Defect domains may offer technological applications associated with the
confinement and transport of charge, as we will investigate in future work.

We acknowledge useful discussions with James Bernard, David Wood,
and David Wu.
LDC was supported by the National Science Foundation under
Grant PHY-0547845 as part of the NSF Career program.

%\bibliographystyle{prsty}
%\bibliography{graphene}

\begin{thebibliography}{10}

\bibitem{novoselov2005}
K.~S. Novoselov {\it et~al.}, Nature {\bf 438},  197  (2005).

\bibitem{zhangYB2005}
Y.~B. Zhang, Y.~W. Tan, H.~L. Stormer, and P. Kim, Nature {\bf 438},  201
  (2005).

\bibitem{geim2007}
A.~K. Geim and K.~S. Novoselov, Nature Materials {\bf 6},  183  (2007).

\bibitem{schedin2006}
F. Schedin {\it et~al.}, e-print cond-mat/0610809 (2006).

\bibitem{silvestrov2007}
P.~G. Silvestrov and K.~B. Efetov, Phys. Rev. Lett. {\bf 98},  016802
  (2007).

\bibitem{green2007}
K. Greene, Technology Review  (2007).

\bibitem{terrones2000}
H. Terrones {\it et~al.}, Phys. Rev. Lett. {\bf 84},  1716  (2000).

\bibitem{fernandezJ2007}
J. Fern\'{a}ndez-Rossier and J.~J. Palacios, Phys. Rev. Lett. {\bf 99},
  177204  (2007).

\bibitem{varchonF2007}
F. Varchon {\it et~al.}, Phys. Rev. Lett. {\bf 99},  126805  (2007).

\bibitem{martins2007}
T.~B. Martins, R.~H. Miwa, A.~J.~R. da~Silva, and A. Fazzio, Phys. Rev. Lett. {\bf 98},  196803  (2007).

\bibitem{chandrasekera2007}
K, Chandrasekera and S, Mukherjee, Comp. Mat. Sci. {\bf 40}, 147 (2007).

\bibitem{borstnika2005}
U. Borštnika, M. Hodošceka, D. Janežica and I. Lukovitsb, J. Chem. Phys. {\bf 411}, 384 (2005).

\bibitem{hamm2007}
M.~T. Lusk and N. Hamm, Phys. Rev. B {\bf 76},  125422  (2007).

\bibitem{hashimoto_2004}
A. Hashimoto {\it et~al.}, Nature {\bf 430},  870  (2004).

\bibitem{lu_2004}
X. Lu, Z. Chen, and P. Schleyer, J. Am. Chem. Soc. {\bf
  127},  20  (2005).

\bibitem{stone_1986}
A. Stone and D. Wales, Chem. Phys. Lett. {\bf 128},  501  (1986).

\bibitem{krasheninnikov2004}
A. Krasheninnikov {\it et~al.}, Carbon {\bf 42},  1021  (2004).

\bibitem{elbarbary2003}
A.~A. El-Barbary {\it et~al.}, Phys. Rev. B {\bf 68},  144107  (2003).

\bibitem{lee_2006}
G.-D. Lee {\it et~al.}, Phys. Rev. B {\bf 74},  245411  (2006).

\bibitem{mpourmpakis_2006}
G. Mpourmpakis, G.~E. Froudakis, and E. Tylianakis, App. Phys. Lett.
  {\bf 89},  233125  (2006).

\bibitem{JChemPhys.113.7756}
B. Delley, J. Chem. Phys. {\bf 113},  7756  (2000).

\bibitem{PhysRevB.45.13244}
J.~P. Perdew and Y. Wang, Phys. Rev. B {\bf 45},  13244  (1992).

\bibitem{terrones2004}
X. Rocquefelte {\it et~al.}, Nano Lett. {\bf 4},  805  (2004).

\bibitem{li:184109}
L. Li, S. Reich, and J. Robertson, Phys. Rev. B {\bf 72},  184109  (2005).

\bibitem{PhysRevB.7.4527}
A.~A. Ahmadieh and H.~A. Rafizadeh, Phys. Rev. B {\bf 7},  4527  (1973).

\bibitem{wilson1955}
E. Wilson, J. Decius, and P. Cross, {\em Molecular Vibrations} (Dover, New
  York, NY, 1955).

\bibitem{rodriguez2004}
J. A. Rodr\'{i}guez-Manzo, F. L\'{o}pez-Urias, M. Terrones, and H. Terrones, Nano
  Lett. {\bf 4},  2179  (2004).

\bibitem{thrower1978}
P.~A. Thrower and R.~M. Mayer, Physica Status Solidi A {\bf 47},  11  (1978).

\bibitem{govind2003}
N. Govind {\it et~al.}, Comp. Mat. Sci. {\bf 28},  250  (2003).

\bibitem{croninAD2006}
A.~D. Cronin and B. McMorran, Phys. Rev. A {\bf 74},  061602(R)  (2006).



\end{thebibliography}

\end{document}